\documentclass[superscriptaddress,twocolumn,amsmath,amssymb,nofootinbib,longbibliography]{revtex4-1}

\usepackage{graphicx}
\usepackage{amsmath}

\usepackage{color}

\usepackage[utf8]{inputenc}

\newcommand{\ba}{\begin{eqnarray}}
\newcommand{\ea}{\end{eqnarray}}

\newcommand{\eps}{\varepsilon}

\newcommand{\tens}[1]{{\ensuremath{\boldsymbol{#1}}}}       
\newcommand{\ts}[1]{{\boldsymbol{#1}}}         

\newcommand{\coder}{{\tens{\delta}}}           
\newcommand{\covd}{{\tens{\nabla}}}            
\newcommand{\pa}{\partial}                     

\newcommand{\A}[1]{A^{\!(#1)}}                 

\newcommand{\be}{\begin{equation}}             
\newcommand{\ee}{\end{equation}}               

\renewcommand{\div}{\ensuremath{\tens d}}

\newcommand{\fieldstrength}{\ensuremath{\mathcal{F}}}
\newcommand{\potential}{\ensuremath{P}}

\begin{document}

\title{Principal Tensor Strikes Again: Separability of Vector Equations with Torsion}

\author{Ramiro Cayuso}
\email{rcayuso@perimeterinstitute.ca}
\affiliation{Perimeter Institute, 31 Caroline Street North, Waterloo, ON, N2L 2Y5, Canada}
\affiliation{Department of Physics and Astronomy, University of Waterloo,
Waterloo, Ontario, Canada, N2L 3G1}

\author{Finnian Gray}
\email{fgray@perimeterinstitute.ca}
\affiliation{Perimeter Institute, 31 Caroline Street North, Waterloo, ON, N2L 2Y5, Canada}
\affiliation{Department of Physics and Astronomy, University of Waterloo,
Waterloo, Ontario, Canada, N2L 3G1}

\author{David Kubiz\v n\'ak}
\email{dkubiznak@perimeterinstitute.ca}
\affiliation{Perimeter Institute, 31 Caroline Street North, Waterloo, ON, N2L 2Y5, Canada}
\affiliation{Department of Physics and Astronomy, University of Waterloo,
Waterloo, Ontario, Canada, N2L 3G1}

\author{Aoibheann Margalit}
\email{amargalit@perimeterinstitute.ca}
\affiliation{Perimeter Institute, 31 Caroline Street North, Waterloo, ON, N2L 2Y5, Canada}
\affiliation{Department of Physics and Astronomy, University of Waterloo,
Waterloo, Ontario, Canada, N2L 3G1}

\author{Renato Gomes Souza}
\email{rsouza@perimeterinstitute.ca}
\affiliation{Perimeter Institute, 31 Caroline Street North, Waterloo, ON, N2L 2Y5, Canada}
\affiliation{Department of Physics and Astronomy, University of Waterloo,
Waterloo, Ontario, Canada, N2L 3G1}
\affiliation{ICTP South American Institute for Fundamental Research, IFT-UNESP, Sao Paulo, SP, Brazil, 01140-070}

\author{Leander Thiele}
\email{lthiele@perimeterinstitute.ca}
\affiliation{Perimeter Institute, 31 Caroline Street North, Waterloo, ON, N2L 2Y5, Canada}
\affiliation{Department of Physics and Astronomy, University of Waterloo,
Waterloo, Ontario, Canada, N2L 3G1}

\date{June 24, 2019}            

\begin{abstract}
Many black hole spacetimes with a 3-form field exhibit a hidden symmetry encoded in a torsion generalization of the principal Killing--Yano tensor. This tensor determines basic properties of such black holes while also underlying the separability of the Hamilton--Jacobi,
Klein--Gordon, and (torsion-modified) Dirac field equations in their background. As a specific example, we consider the Chong--Cveti{\v c}--L{\" u}--Pope black hole of $D=5$ minimal gauged
supergravity and show that the torsion-modified vector field equations can also be separated, with the principal tensor playing a key role in the separability ansatz. For comparison, separability of the Proca field in higher-dimensional Kerr--NUT--AdS spacetimes (including new explicit formulae in odd dimensions) is also presented.
\end{abstract}

\maketitle

\section{Introduction}
Hidden symmetries of dynamics play a crucial role for the study of diverse physical systems that can be relativistic or not, with or without gravity, classical or quantum \cite{Cariglia:2014ysa}. Particularly useful for black hole physics is the hidden symmetry of the {\em principal Killing--Yano tensor} \cite{Frolov:2017kze}.
Such a tensor not only generates other explicit and hidden symmetries, it also determines the algebraic type of the solution, and allows for a separation of variables when studying physical field equations in black hole spacetimes.

The Golden Era of field equation separability in Kerr geometry spanned the late '60s and '70s.
During this period, separable solutions were found for particles following geodesic motion, test scalar and spinor fields, as well as massless vector (spin 1) and tensor (spin 2) fields \cite{Carter:1968cmp, Carter:1968rr, Unruh:1973, Chandrasekhar:1976, Teukolsky:1972, Teukolsky:1973}.
Much later, after the generalization of Kerr geometry to higher dimensions by Myers and Perry \cite{MyersPerry:1986}, a flurry of papers extended the results for geodesic motion \cite{Page:2006ka}, and test scalar \cite{Frolov:2007cb} and spinor \cite{OotaYasui:2008} fields to arbitrary dimensions---all thanks to the principal tensor. However, the appropriate separation scheme for vector and tensor in dimensions $D>4$ remained elusive.

A breakthrough on this front came in 2017 when Lunin demonstrated the separability of Maxwell's equations in Myers--Perry-(A)dS geometry \cite{Gibbons:2004uw}. Lunin's approach was novel in that it provided a separable ansatz for the vector potential rather than the field strength,
a method that had previously seen success in $D=4$ dimensions \cite{Teukolsky:1972, Teukolsky:1973}.
In 2018, Frolov, Krtous, and Kubiznak showed that Lunin's ansatz can be written in a covariant form, in terms of the principal tensor \cite{Frolov:2018pys, Krtous:2018bvk}, allowing them to extend Lunin's result to general (possibly off-shell) Kerr--NUT--AdS spacetimes \cite{Chen:2006xh}. The separation of massive vector (\emph{Proca}) field perturbations in these spacetimes (an achievement previously absent even for the four-dimensional Kerr geometry) followed shortly after that \cite{Frolov:2018ezx}, see also \cite{Dolan:2018dqv, Frolov:2018eza,Dolan:2019hcw}.

The separability of the vector field hinges on the existence of the principal tensor. Such a tensor: i) determines the canonical (preferred) coordinates in which the separation occurs, ii) generates the towers of explicit and hidden symmetries linked to the `symmetry operators' of the separated vector equation, and iii) explicitly enters the separation ansatz for the vector potential $\tens{P}$. Namely, this ansatz can be written in the following covariant form:
\be\label{LFKK}
P^a=B^{ab}\nabla_b Z\,,
\ee
where the tensor $\tens{B}$ is determined by the principal tensor $\tens{h}$ and the metric $\tens{g}$ according to
\be\label{Btens}
B^{ab}(g_{bc}+i\mu h_{bc})=\delta^a_c\,,
\ee
with $\mu$ a `separation constant'.
The solution for the scalar functions $Z$ is then sought in a standard multiplicative separable form. In what follows, we shall refer to the ansatz
\eqref{LFKK} with $\tens{B}$ given by \eqref{Btens} as the {\em Lunin--Frolov--Krtous--Kubiznak (LFKK) ansatz}. Remarkably, the LFKK ansatz works equally well for both massless and massive vector perturbations.
It is also valid in \emph{any} dimension, even or odd, as discussed in Appendix \ref{appendix}.

However striking the above results are, they have one serious drawback: the existence of the principal tensor is largely limited to vacuum black hole spacetimes. In fact, the conditions for its existence are so strong that they uniquely determine a (rather restricted though physically interesting) class of admissible
spacetimes---the off shell Kerr--NUT--AdS metrics \cite{Krtous:2008tb}.  For this reason various
generalizations of the notion of hidden symmetries, that would allow for more general spacetimes
while preserving integrability features of the symmetry, were sought. One such generalization, that
of the Killing--Yano tensor with torsion \cite{yano1953curvature, Rietdijk:1995ye, Kubiznak:2009qi}, turns out to be quite fruitful.
Such a symmetry exists in a number of supergravity and string theory spacetimes, where the torsion
can be naturally identified with a defining 3-form of the theory. Although less restrictive, the principal
tensor with torsion still implies the essential integrability features of its torsion-less cousin and underlies separability of Hamilton--Jacobi, Klein--Gordon, and torsion-modified Dirac equations on the black hole background.

With this in mind a natural question arises: is the vector field separability described above limited to the vacuum spacetimes? It is the purpose of the present paper to show that it is not so. To this end,
we concentrate on a prototype non-vacuum black hole spacetime known to admit the principal tensor with torsion \cite{Kubiznak:2009qi},
the Chong--Cveti\v{c}--L\"u--Pope black hole \cite{Chong:2005hr}
of $D=5$ minimal gauged supergravity. We show that the LFKK ansatz can be used for separability
of the (properly torsion modified) vector equation in this background. Our  result shows that,
despite the presence of torsion considerably weakening the structure, the principal tensor with
torsion remains a powerful tool for separability. At the
same time it means that the applicability of the LFKK ansatz goes far beyond the original setup in
which it was first considered.

Our paper is organized as follows. In Sec. \ref{KYTT}, we review the torsion generalization of Killing--Yano tensors and introduce the principal tensor with torsion.
Sec.~\ref{BH} is devoted to the Chong--Cveti\v{c}--L\"u--Pope black hole of the $D=5$ minimal gauged supergravity and its basic characteristics. The novel contribution of the paper comes in Sec.~\ref{Sep} where the torsion modified massive vector (Troca) equation is introduced and shown to separate using the LFKK separability ansatz. We conclude in Sec.~\ref{summary}. Appendix~\ref{appendix} gathers the results on separability of the vector equations in the off-shell Kerr--NUT--AdS spacetimes---the results in odd dimensions are presented for the first time and allow us to compare the five-dimensional vacuum result to the supergravity case studied in the main text.

\section{Killing--Yano tensors with torsion}\label{KYTT}
Let us start by briefly  recapitulating  a `torsion generalization' of Killing--Yano tensors which has applications for a variety of supergravity black hole solutions.

In what follows we assume that the torsion is completely antisymmetric and described by a 3-form $\tens{T}$, with the  standard torsion tensor given by  $T^{d}_{ab}=T_{abc}g^{cd}$, where $\tens{g}$ is the metric. The torsion connection $\tens{\nabla}^T$ acting on a vector field $\ts{X}$ is defined as
\be\label{j1-m}
\nabla^T_{\!a} X^b = \nabla_{\!a} X^b + \frac{1}{2}\,T_{ac}^b X^c\,,
\ee
where $\tens{\nabla}$ is the Levi-Civita (torsion-free) connection. The connection $\covd^T$ satisfies the metricity condition, $\covd^T \tens{g}=0$, and has the same geodesics as $\covd$. It induces a
connection acting on forms. Namely, let $\tens{\omega}$ be a $p$-form, then
\be\label{j5-m}
\covd^T_{\!\ts{X}} \tens{\omega}=\covd_{\!\ts{X}} \tens{\omega}
  -\frac{1}{2} \bigl(\ts{X}\cdot\tens{T}\bigr)\underset{1}{\wedge} \tens{\omega}\,,
\ee
where we have used a notation of contracted wedge product introduced in \cite{Houri:2010qc}, defined for a $p$-form $\tens{\alpha}$ and $q$-form $\tens{\beta}$ as
\ba\label{ContrProd}
(\alpha\underset{m}{\wedge}\beta)_{a_1\dots a_{p{-}m}b_1\dots b_{q{-}m}}&&\nonumber\\
=\frac{(p+q-2m)!}{(p\!-\!m)!(q\!-\!m)!}&&\!
\alpha_{c_1\dots c_m[a_1\dots a_{p{-}m}}\beta^{c_1\dots c_m}{}_{b_1\dots b_{q{-}m}]}.\quad\
\ea
Equipped with this, one can define the following two operations:
\begin{align}
\tens{d}^T \tens{\omega}&\equiv \covd^T \wedge \tens{\omega}=\tens{d\omega}-\tens{T}\underset{1}{\wedge}\tens{\omega}\,,
\label{j6-m}\\
\coder^T \tens{\omega}&\equiv-\covd^T\cdot\tens{\omega}=\coder\tens{\omega}-\frac{1}{2}\,\tens{T}\underset{2}{\wedge} \tens{\omega}\,.\label{j7-m}
\end{align}
Note that in particular we have $\coder^T \tens{T}=\coder \tens{T}$.

A {\em conformal Killing--Yano tensor with torsion} $\tens{k}$  is a $p$-form which for any
vector field $\tens{X}$ satisfies the following equation \cite{yano1953curvature, Rietdijk:1995ye, Kubiznak:2009qi}:
\be\label{generalizedCKY}
\nabla^T_X \tens{k}-\frac{1}{p+1}\tens{X}\cdot \tens{d}^T \tens{k}+\frac{1}{D-p+1}
\tens{X} \wedge \tens{\delta}^T \tens{k}=0\,,
\ee
where $D$ stands for the total number of spacetime dimensions.
In analogy with the Killing--Yano tensors defined with respect to the Levi-Civita connection,
a conformal Killing--Yano tensor with torsion $\tens{f}$ obeying
$\tens{\delta}^T \tens{f}=0$ is called a {\em Killing--Yano tensor with torsion}, and a  conformal Killing--Yano tensor with torsion $\tens{h}$ obeying
$\tens{d}^T \tens{h}=0$ is a {\em closed conformal Killing--Yano tensor with torsion}.

Despite the presence of torsion, the conformal Killing--Yano tensors with torsion possess many remarkable properties.
The following three are especially important for `generating other symmetries'  and separability of test field equations (see \cite{Kubiznak:2009qi,Houri:2010fr} for the proof and other properties):
\begin{enumerate}
\item The Hodge star $\tens{*}$ maps  conformal Killing--Yano with torsion $p$-forms to conformal Killing--Yano with torsion $(D-p)$-forms. In particular, the Hodge star of a closed conformal Killing--Yano with torsion $p$-form is a Killing--Yano with torsion $(D-p)$-form and vice versa.
\item Closed conformal Killing--Yano tensors with torsion form a (graded) algebra with respect to a wedge product. Namely, let $\tens{h}_1$ and $\tens{h}_2$ be a closed conformal Killing--Yano tensor with torsion $p$-form and $q$-form, respectively, then $\tens{h}_3=\tens{h}_1 \wedge \tens{h}_2$ is a closed conformal Killing--Yano with torsion $(p+q)$-form.
\item Let $\tens{h}$ and $\tens{k}$ be two (conformal) Killing--Yano tensors with torsion of rank $p$. Then
\be
K_{ab}=h_{(a |c_1\ldots c_{p-1}|}k_{b)}{}^{c_1\ldots c_{p-1}}
\ee
is a (conformal) Killing tensor of rank 2.
\end{enumerate}

In what follows, we shall concentrate on a {\em principal tensor} with torsion, $\tens{h}$, which is a non-degenerate closed conformal Killing--Yano with torsion 2-form. It obeys
\be\label{principal}
\nabla^T_X \tens{h}=\tens{X} \wedge \tens{\xi}\,,\quad \tens{\xi}=-\frac{1}{D-p+1} {\delta}^T \tens{h}\,.
\ee
The condition of non-degeneracy means that $\tens{h}$ has the maximal possible (matrix) rank and possesses the maximal number
of functionally independent eigenvalues.

Starting from one such tensor, one can generate (via the three properties above) the towers of Killing tensors, and (closed conformal) Killing--Yano tensors with torsion. In their turn, such symmetries can typically be associated with symmetry operators for a given field operator. For example,
Killing tensors give rise to operators commuting with the scalar wave operator, and (closed conformal) Killing--Yano tensors with torsion to operators commuting with the torsion Dirac operator. When a full set of mutually commuting operators can be found, one can typically separate the corresponding field equation. It is in this respect
the principal tensor underlies separability of various field equations.

The existence of the principal tensor imposes severe restrictions on the spacetime. In the torsion-less case, such restrictions uniquely determine the Kerr--NUT--AdS class of black hole spacetimes \cite{Krtous:2008tb} (see also \cite{Frolov:2017whj}). Although no full classification is available for spacetimes with torsion, nor is it clear if such spacetimes have to admit any isometries \cite{Houri:2012eq}, several explicit examples of supergravity solutions with a principal tensor
where the torsion is naturally identified with a 3-form field strength of the theory are known. Among them, perhaps the most `beautiful' are
the $D$-dimensional Kerr--Sen spacetimes  \cite{Houri:2010fr} and black holes of $D=5$  minimal gauged supergravity \cite{Kubiznak:2009qi}.

In this paper we concentrate on the latter, very `clean' solution without scalar fields, known as the Chong--Cveti{\v c}--L{\" u}--Pope black hole \cite{Chong:2005hr}. It is known that for such a solution the principal tensor guarantees integrability of the geodesic motion \cite{Davis:2005ys}, as well as separability of scalar \cite{Davis:2005ys} and modified Dirac \cite{Davis:2005ys, Wu:2009cn, Wu:2009ug} equations. Our aim is to show that it also guarantees the separability of properly torsion modified (massive) vector field equations.

\section{Black hole of minimal gauged supergravity}\label{BH}

The bosonic sector of $D=5$ minimal gauged supergravity is governed by the Lagrangian
\be\label{L}
\pounds=\tens{*}(R+\Lambda)-\frac{1}{2}\tens{F}\wedge \tens{*F}\!+
\frac{1}{3\sqrt{3}}\,\tens{F} \wedge \tens{F}\wedge \tens{A}\,,
\ee
where $\Lambda$ is the cosmological constant. This yields the following set of Maxwell and Einstein equations:
\ba
\tens{dF}=0\,,\quad \tens{d* F}-\frac{1}{\sqrt{3}}\,\tens{F}\wedge\tens{F}\!&=&\!0\,,\quad
\label{F}\label{Maxwell}\\
R_{ab}-\frac{1}{2}\Bigl(F_{ac}F_b^{\ c}-\frac{1}{6}\,g_{ab}F^2\Bigr)+\frac{1}{3}\Lambda g_{ab}\!&=&\!0\,.
\label{Einstein}
\ea
In this case the torsion can be identified with the Maxwell field strength according to \cite{Kubiznak:2009qi}
\be\label{TF}
\tens{T}=-\frac{1}{\sqrt{3}}\, \tens{*F}\,.
\ee
Having done so, the Maxwell equations can be written as
\be
\tens{\delta}^T\tens{T}=0\,, \quad \tens{d}^T\tens{T}=0\,.
\ee
In other words, the torsion $\tens{T}$ is `$T$-harmonic'.
Here, the first equality follows from the fact that $\coder^T \tens{T}=\coder \tens{T}$, while the second is related to the special property
in five dimensions (with Lorentzian signature),
\be
\tens{d}^T \tens{\omega}=\tens{d\omega}+(\tens{*T})\wedge (\tens{*\omega})\,,
\ee
valid for any 3-form $\tens{\omega}$. The principal tensor equation \eqref{principal} can now explicitly be written as
\ba\label{PCKY2}
\nabla_c h_{ab}&=&2g_{c[a}\xi_{b]}+\frac{1}{\sqrt{3}}\,(*F)_{cd[a}h^d_{\ \,b]}\,,\nonumber\\
\xi^a&=&\frac{1}{4} \nabla_b h^{ba}-\frac{1}{2\sqrt{3}}(*F)^{abc} h_{bc}\,.
\ea

A general doubly spinning charged black hole solution in this theory has been constructed by
Chong, Cveti\v{c}, L\"u, and Pope \cite{Chong:2005hr}. It can be written in a symmetric Wick-rotated form, c.f. \cite{Kubiznak:2009qi}:
\ba
\tens{g}\!&=&\!\sum_{\mu=1,2}\bigl(\tens{\omega}^{\mu}\tens{\omega}^{\mu}+
\tens{\tilde \omega}^{\mu}\tens{\tilde\omega}^{\mu}\bigr)
+\tens{\omega}^{0}\tens{\omega}^{0}\,,\label{can_odd}\\
\tens{A}\!&=&\!\sqrt{3}c(\tens{A}_1+\tens{A}_2)\,,\label{A}
\ea
where
\ba\label{omega}
\tens{\omega}^{1} \!&=&\sqrt{\frac{U_1}{X_1}}\,\tens{d}x_1\,,\quad
\tens{\tilde \omega}^{1}=\sqrt{\frac{X_1}{U_1}}(\tens{d}\psi_0+x_2^2\tens{d}\psi_1)\,,\nonumber\\
\tens{\omega}^{2} \!&=&\!\sqrt{\frac{U_2}{X_2}}\,\tens{d}x_2\,,\quad
\tens{\tilde \omega}^{2}=\sqrt{\frac{X_2}{U_2}}(\tens{d}\psi_0+x_1^2\tens{d}\psi_1)\,,\nonumber\\
\tens{\omega}^{0}\!&=&\!\frac{ic}{x_1x_2}\!\Bigl[\tens{d}\psi_0\!+\!(x_1^2\!+\!x_2^2)\tens{d}\psi_1\!+\!x_1^2x_2^2\tens{d}\psi_2\!
-\!x_2^2\tens{A}_1\!-\!x_1^2\tens{A}_2\Bigr]\,,\nonumber\\
\tens{A}_1\!&=&\!-\frac{e_1}{U_1}(\tens{d}\psi_0+x_2^2\tens{d}\psi_2)\,,\quad \tens{A}_2=-\frac{e_2}{U_2}(\tens{d}\psi_0+x_1^2\tens{d}\psi_1)\,,\nonumber\\
U_1&=&x_2^2-x_1^2=-U_2\,.
\ea
The  solution is stationary and axisymmetric, corresponding to three Killing vectors $\pa_{\psi_0}, \pa_{\psi_1}, \pa_{\psi_2}$,  and possesses two non-trivial coordinates $x_1$ and $x_2$. Here we choose $x_2>x_1>0$ and note that the metric written in this form has $\det {g}<0$. We have also used a `symmetric gauge' for the $U(1)$ potential; the electric charge of the Maxwell field $\ts{F}=\ts{dA}$ depends on a difference $(e_1-e_2)$.

In order to satisfy the Einstein--Maxwell equations, the metric functions take the following form:
\ba
X_1&=&A+Cx_1^2-\frac{\Lambda}{12}x_1^4+\frac{c^2(1+e_1)^2}{x_1^2}\,,\nonumber\\
X_2&=&B+Cx_2^2-\frac{\Lambda}{12}x_2^4+\frac{c^2(1+e_2)^2}{x_2^2}\,\label{XY},
\ea
where of the four free parameters $A,B,C, c$ only three are physical (one can be scaled away) and are related to the mass and two rotations. As per usual, the separability property shown below does not need the special form \eqref{XY} and occurs ``off-shell'', for arbitrary functions
\be
X_1=X_1(x_1)\,,\quad X_2=X_2(x_2)\,.
\ee

As shown in \cite{Kubiznak:2009qi}, the spacetime admits a principal tensor with torsion, which
takes the form
\be
\ts{h}=\sum_{\mu=1,2} x_\mu \tens{\omega}^{\mu}\wedge \tens{\tilde \omega}^{\mu}\,.
\ee
Interestingly, the torsion \eqref{TF} in Chong--Cveti{\v c}--L{\"u}--Pope spacetimes is very special as it satisfies the following conditions:
\be
(*F)_{d[ab}h^d_{\ \,c]}=0\,,\quad (*F)_{abc} h^{bc}=0\,.
\ee
This implies that the tensor is not only $\tens{d}^T$-closed (as it must be), but it is also $\tens{d}$-closed and obeys:
\be
\tens{d}^T\tens{h}=\tens{dh}=0\,,\quad \tens{\xi}=-\frac{1}{4}\tens{\delta}^T \tens{h}=-\frac{1}{4}\tens{\delta} \tens{h}=\pa_{\psi_0}\,.
\ee
Therefore it can be locally written in terms of a potential
\be
\tens{h}=\tens{db}\,,\quad \tens{b}=-\frac{1}{2}\Bigl[(x_1^2+x_2^2)\tens{d}\psi_0+x_1^2x_2^2 \tens{d}\psi_1\Bigr]\,.
\ee

Using the properties of closed conformal Killing--Yano tensors with torsion, the principal tensor generates a Killing--Yano with torsion 3-form $\tens{*h}$, and a rank-2 Killing tensor
\be\label{KT}
K_{ab}=(*h)_{acd}(*h)_b^{\ cd}=h_{ac}h_b^{\ c}-\frac{1}{2}g_{ab}h^2\,.
\ee
Such symmetries are responsible for separability of the Hamilton--Jacobi, Klein--Gordon, and torsion-modified Dirac equations in these spacetimes.

\section{Separability of vector perturbations}\label{Sep}

\subsection{Troca equation}
Let us now proceed and consider a test massive vector field \tens{\potential} on the above background. It is reasonable to expect that,
similar to the Dirac case \cite{Kubiznak:2009qi,Houri:2010qc}, the corresponding Proca equation will pick up the due torsion generalization.
In what follows we shall argue that the natural sourceless massive vector equation to consider is
\be
\label{eq:troca}
\covd \cdot \tens{\fieldstrength} - m^2\tens{P} = 0\,,
\ee
where $m$ is the mass of the field, and the field strength $\tens\fieldstrength$ is defined via the torsion exterior derivative,
\be
\tens\fieldstrength \equiv \tens{d}^T\!\tens{P}=\div\tens{P} - \tens{P}\cdot \tens T\,.
\ee
Being a torsion generalization of the Proca equation, we shall refer to the equation \eqref{eq:troca} as a `{\em Troca equation}'.
It implies the `Lorenz condition'
\be\label{eq:Lorenz}
\covd \cdot \tens{P}=0\,.
\ee

To motivate the above form of the Troca equation, we demand that it is linear in $\tens{P}$, reduces to the Proca equation in the absence of torsion, and would obey the current conservation in the presence of sources. We have three natural candidates for generalizing the `Maxwell operator' $\covd\cdot \tens{dP}$, namely:
\be
\tens{O}_1=\covd \cdot \tens{d}^T\!\tens{P}\,,\quad
\tens{O}_2=\covd^T \cdot \tens{d}\!\tens{P}\,,\quad
\tens{O}_3=\covd^T \cdot \tens{d}^T\!\tens{P}\,.
\ee
However, the last two do not obey the current conservation equation. Indeed,
due to $\covd^T \cdot (\covd^T \cdot \ )\neq  0$, we have
$\covd\cdot \tens{O}_2=\covd^T \cdot \tens{O}_2\neq 0$, and similarly for $\tens{O}_3$. So we are left with $\tens{O}_1$ which, when extended to the massive case, yields the Troca equation \eqref{eq:troca}.

Let us also note that the choice of operator $\tens{O}_1$  is `consistent' with the Maxwell equation for the background Maxwell field. Namely, due to the identity
\be
\tens{X}\cdot \tens{*\omega}=\tens{*}(\tens{\omega}\wedge \tens{X})\,,
\ee
valid for any vector $\tens{X}$ and any $p$-form $\tens{\omega}$, the field equations \eqref{Maxwell} can be written as
\ba
0&=&\tens{d* F}-\frac{1}{\sqrt{3}}\,\tens{F}\wedge\tens{F}=\tens{d*dA}-\tens{d}\Bigl(\frac{1}{\sqrt{3}}\tens{F}\wedge \tens{A}\Bigr)\nonumber\\
&=&\tens{d*dA}+\tens{d*}\Bigl(\tens{A}\cdot \frac{1}{\sqrt{3}}\tens{*F}\Bigr)\nonumber\\
&=&\tens{d*}\tens{d}^T\tens{A}\,.
\ea
That is, identifying $\tens{A}$ with the Proca field in the test field approximation 
\be
\covd\cdot \tens{d}^T\tens{P}=0\,,
\ee
which is the massless Troca equation \eqref{eq:troca} (upon treating the torsion as an independent field).

\subsection{Separability}
Having identified the Troca equation \eqref{eq:troca}, let us now find its general solution in the supergravity background \eqref{can_odd}. To this purpose we
employ the LFKK ansatz:
\be\label{LFKK2}
P^a=B^{ab}\nabla_b Z\,,\quad B^{ab}(g_{bc}+i\mu h_{bc})=\delta^a_c\,,
\ee
and seek the solution in a separated form
\begin{equation}\label{separ}
Z = {\underset{\nu = 1,2}{\prod}} R_{\nu} \left( x_{\nu} \right) \exp \Bigl[i \overset{2}{\underset{j = 0}{\sum}} L_{j} \psi_{j} \Bigr]\,,
\end{equation}
where $\{\mu, L_0, L_1, L_2\}$ are four `separation constants'.

As in the case without torsion, it is useful to start with the Lorenz condition \eqref{eq:Lorenz}. We find
\be\label{eq: ODEs for mode functions}
\nabla_a P^a=
Z\sum_{\nu=1,2}\frac{1}{U_{\nu}}\frac{\mathcal{D}_{\nu}R_{\nu}}{q_{\nu} R_{\nu}}\,,
\ee
where the differential operator $\mathcal{D}_{\nu}$ is given by
\ba\label{Operator Sugra}
\mathcal{D}_{\nu} &=& \dfrac{q_{\nu}}{x_{\nu}}\partial_{\nu}\left( \frac{X_{\nu} x_{\nu}}{q_{\nu}}\partial_{\nu} \right) - \frac{1}{X_{\nu}} \Bigl(\overset{2}{\underset{j=0}{\sum}} (-x_{\nu}^2)^{1-j}{L}_{j\nu}\Bigr)^{\!2} \nonumber \\
&&+   \frac{ 2 \mu }{q_{\nu}}\ \overset{2}{\underset{j=0}{\sum}} (-\mu^2)^{j-1}{L}_{j\nu}  + \frac{L_2^2 q_{\nu}}{c^2 x_{\nu}^2 }\,,
\ea
and we have defined
\be
q_\nu=1-\mu^2 x_\nu^2\,, \quad {L}_{j\nu} = L_{j}(1+\delta_{j2}e_{\nu})\,,
\ee
the latter definition being essentially the only difference when compared to the five-dimensional torsion-less case, c.f. \eqref{eq: Operator odd D}.

In order to impose the Lorenz condition, we could now follow the procedure developed in \cite{Frolov:2018ezx}. Instead, let us proceed in a slightly different way, by using the following `{separability lemma}' \cite{Frolov:2007cb}:\\
{\bf Lemma.}
{\em The most general solution of
\be
\sum_{\nu=1}^n \frac{f_\nu(x_\nu)}{U_\nu}=0\,\quad \mbox{where}\quad U_{\nu} = \overset{n}{\underset{\mu \neq \nu}{\underset{\mu = 1}{\prod}}} \left( x_{\mu}^2 - x_{\nu}^2 \right)\,,
\ee
is given by
\be\label{sepfnu}
f_\nu=\sum_{j=0}^{n-2} C_j(-x_\nu^2)^{n-2-j}\,,
\ee
where $C_j$ are arbitrary constants.}

Thus, demanding  $\nabla_a P^a=0$, using the expression \eqref{eq: ODEs for mode functions}, and the above lemma for $n=2$ and $f_\nu=\mathcal{D}_{\nu}R_{\nu}/(q_{\nu} R_{\nu})$, yields the separated equations
\be\label{eq: separation equations0}
\mathcal{D}_{\nu}R_{\nu}=q_{\nu}f_\nu R_{\nu}\,,
\ee
where $f_\nu$ is given by \eqref{sepfnu}, that is, $f_\nu=C_0$.

With this at hand, let us now turn to the Troca equation \eqref{eq:troca}.
Using the ansatz \eqref{LFKK2} and the Lorenz condition \eqref{eq:Lorenz}, the L.H.S. of the Troca equation takes the following form:
\begin{equation}
    \nabla_b \fieldstrength^{ba} - m^2 \potential^a =  B^{ab} \nabla_{b} J\,,
\end{equation}
where the `current' $J$ is, similar to the Kerr--NUT--AdS case~\cite{Frolov:2018ezx}, given by
\begin{equation}
    J = \Box Z - 2 i \mu \xi_{a} B^{ab} \partial_{b}Z - m^2 Z\,,
\end{equation}
or more explicitly,
\begin{equation}\label{eq: current J}
    J = Z \sum_{\nu=1,2}\frac{1}{U_{\nu}R_{\nu}}\Bigl[ \mathcal{D}_{\nu} - m^2 \left( -x_{\nu}^2 \right) \Bigr]R_{\nu}\,.
\end{equation}
Using the separation equations \eqref{eq: separation equations0} and the lemma again, we find
\be
J=-Z(\mu^2 C_0-m^2)\sum_{\nu=1,2}\frac{x_\nu^2}{U_\nu}=Z(\mu^2 C_0-m^2)\,.
\ee
In order for this to vanish, we require
\be
C_0=\frac{m^2}{\mu^2}\,.
\ee
Of course, in the case of massless vectors, we can set $m=0$.

To summarize, we have shown  the separation of variables for the Troca equation \eqref{eq:troca} in the Chong--Cveti\v{c}--L\"u--Pope black hole spacetime. The solution can be found in the form of the LFKK ansatz \eqref{LFKK2}, where the scalar function $Z$ is written in the multiplicative separated form \eqref{separ}, and the modes  $R_\nu$ satisfy the ordinary differential equations \eqref{eq: separation equations0} with $f_\nu=C_0=m^2/\mu^2$. The obtained solution is general in that it
depends on four independent separation constants $\{\mu, L_0, L_1, L_2\}$. It remains to be seen if, similar to the Kerr--NUT--AdS case \cite{Dolan:2018dqv}, all polarizations (four in the case of massive field and three for $m=0$) are captured by our solution.

\section{Conclusions}\label{summary}

The principal tensor is a very powerful object. It uniquely characterizes the class of vacuum black hole spacetimes (known as the Kerr--NUT--AdS metrics), and it generates towers of explicit and hidden symmetries. In turn, such symmetries underlie separability of the Hamilton--Jacobi, Klein--Gordon, Dirac, and as only recently shown also Maxwell and Proca equations in these spacetimes. The key to separating the vector equations is not to concentrate on the field strength (as previously thought) but rather employ a new LFKK separability ansatz \eqref{LFKK} and \eqref{Btens} for the vector potential itself.

In this paper we have shown that the applicability of the LFKK ansatz goes far beyond the realm previously expected. Namely, we have demonstrated
the separability of the vector field equation in the background of the Chong--Cveti\v{c}--L\"u--Pope black hole of minimal gauged supergravity. Such a black hole
no longer possesses a principal tensor. However, upon identifying the Maxwell 3-form of the theory with torsion, a weaker structure, the principal tensor with torsion, \emph{is} present. Remarkably, such a structure enters the LFKK ansatz in precisely the same way as the standard (vacuum) principal tensor and allows one to separate the naturally torsion modified vector (Troca) field equations: `{\em principal tensor strikes again'}.  This result opens future horizons for applicability of both the LFKK ansatz and the torsion modified principal tensor. It is an interesting open question to see where the principal tensor is going to strike next.

\section*{Acknowledgements}
\label{sc:acknowledgements}
We would like to thank Ma{\"i}t{\'e} Dupuis and Lenka Bojdova for organizing the PSI Winter School where this project was mostly completed and the PSI program for facilitating this research.
The work was supported in part by the Natural Sciences and Engineering Research Council of Canada.
Research at Perimeter Institute is supported in part by the Government of Canada through the Department of Innovation, Science and Economic Development Canada and by the Province of Ontario through the Ministry of Economic Development, Job Creation and Trade.
R.G.S thanks IFT-UNESP/ICTP-SAIFR and CNPq for partial financial support.
L.T. acknowledges support by the Studienstiftung des Deutschen Volkes.

\appendix

\section{Separability of Proca equation in Kerr--NUT--AdS spacetimes}\label{appendix}

The separability of (massive) vector equations in Kerr--NUT--AdS spacetimes has been recently demonstrated in \cite{FrolovEtAl:2018b}. However, although the result was claimed to be true in all dimensions, the explicit formulae were only stated in even dimensions. It is the purpose of this appendix to rectify this situation and present the separated equations in all dimensions, with the necessary alterations for odd-dimensions.
This allows us then to compare the five-dimensional Kerr--NUT--AdS result to the five-dimensional SUGRA case studied in the main text.

To treat the odd $(\epsilon=1)$ and even ($\epsilon=0)$ dimensions simultaneously, we parameterize the total number of spacetime dimensions $D$ as
\be
D=2n+\epsilon\,.
\ee
The (off-shell) Kerr--NUT--AdS spacetime is the most general metric that admit the principal tensor \emph{without} torsion. The metric and the principal tensor
are
\ba
    \tens g &=& \sum_{\nu = 1}^{n} (\tens\omega^{\nu} \tens\omega^{\nu} +
    \tens{\tilde\omega}^{\nu}\tens{\tilde\omega}^{\nu})+\eps \tens\omega^{0} \tens\omega^{0}\,, \label{eq: reference metric}\\
    \tens h &=& \sum_{\nu = 1}^{n} x_{\nu} \tens\omega^{\nu} \wedge \tens{\tilde\omega}^{\nu}\,, \label{eq: principal tensorAA}
   \ea
where the orthonormal frame is given by
\ba\label{eq: orthonormal frame in odd Kerr}
\tens{\omega}^{\nu} &=& \sqrt{\dfrac{U_{\nu}}{X_{\nu}}} \tens{d}x_{\nu}\,,\quad
\tens{\tilde\omega}^{\nu} = \sqrt{\dfrac{X_{\nu}}{U_{\nu}}} \ \overset{n-1}{\underset{j = 0}{\sum}} A^{\left( j \right)}_{\nu} \tens{d} \psi_{j}\,, \label{omega_nu_bar}\nonumber\\
\tens \omega^{0} &=&\dfrac{ic}{\sqrt{A^{\left( n \right)}}} \ \overset{n}{\underset{j = 0}{\sum}} A^{\left( j \right)}\tens{d} \psi_{j}\,, \label{omega_epsilon}
\ea
and the metric functions are given by
\ba\label{AUdefs}
  \A{k}&=&\!\!\!\!\!\sum_{\substack{\nu_1,\dots,\nu_k=1\\\nu_1<\dots<\nu_k}}^n\!\!\!\!\!x^2_{\nu_1}\dots x^2_{\nu_k}\;,
\quad
\A{j}_{\mu}=\!\!\!\!\!\sum_{\substack{\nu_1,\dots,\nu_j=1\\\nu_1<\dots<\nu_j\\\nu_i\ne\mu}}^n\!\!\!\!\!x^2_{\nu_1}\dots x^2_{\nu_j}\;,\nonumber\\
  U_{\mu}&=&\prod_{\substack{\nu=1\\\nu\ne\mu}}^n(x_{\nu}^2-x_{\mu}^2)\;.
\ea
The metric is a solution of Einstein--$\Lambda$ equations, provided we
set
\be\label{Xsol}
X_\mu=-2b_\mu x_\mu^{1-\epsilon}+\sum_{k=\epsilon}^n c_k x_\mu^{2k}+\epsilon \frac{c^2}{x_\mu^2}\,,
\ee
where the constants $c, c_k$, and $b_\mu$ are related to the cosmological constant, angular momenta, mass, and NUT charges \cite{Frolov:2017kze}.
As per usual in these spacetimes, the separability property demonstrated below remains true off-shell, for any functions
\be
X_\mu=X_\mu(x_\mu)\,.
\ee

We are interested in finding a general solution of the Proca equation
\begin{equation}\label{eq: Proca equation}
   \nabla_{a}{F}^{ab} - m^2\potential^{b} = 0\,,
\end{equation}
in the background \eqref{eq: reference metric}, where $\tens{F}=\tens{dP}$. To this purpose we employ the
LFKK ansatz
\begin{equation}\label{LFKK3}
    \potential^{a}=B^{ab} \nabla_{b} Z\,,\quad  B^{ab}(g_{bc}+i\mu h_{bc}) = \delta^a_c\,,
\end{equation}
and seek the solution in the separated form
\begin{equation}\label{sep3}
Z = \overset{n}{\underset{\nu = 1}{\prod}} R_{\nu} \left( x_{\nu} \right) \exp \Bigl(i \overset{n-1+\epsilon}{\underset{j = 0}{\sum}} L_{j} \psi_{j} \Bigr)\,.
\end{equation}

An immediate consequence of the Proca equation \eqref{eq: Proca equation} is the ``Lorenz condition",
\begin{equation}\label{eq: appendix Lorenz condition}
    \nabla_a \potential^a = 0\,.
\end{equation}
This provides an explicit linear ODE for the functions $R_{\nu}(x_{\nu})$ appearing in the separation ansatz \eqref{sep3}, namely,
\begin{equation}\label{eq: appendix ODEs for mode functions}
    0 = \nabla_a \potential^a = Z\sum_{\nu=1}^{n}\frac{1}{U_{\nu}}\frac{\mathcal{D}_{\nu}R_{\nu}}{q_{\nu} R_{\nu}}\,,
\end{equation}
where $q_\nu=1 - \mu^2 x_{\nu}^2$ and the differential operator $\mathcal{D}_{\nu}$ is given by
\ba\label{eq: Operator odd D}
\nonumber \mathcal{D}_{\nu} &=& \dfrac{q_{\nu}}{x_{\nu}^\epsilon}\partial_{\nu}\Bigl[ \frac{X_{\nu} x_{\nu}^\epsilon}{q_{\nu}}\partial_{\nu} \Bigr] - \frac{1}{X_{\nu}} \Bigl[\overset{n-1+\epsilon}{\underset{j=0}{\sum}}\!\! (-x_{\nu}^2)^{N-1-j}L_j\Bigr]^{2} \\
&+&\mu\bigl(\frac{2}{q_\nu}+\epsilon-1\bigr)\overset{n-1+\epsilon}{\underset{j=0}{\sum}} (-\mu^2)^{j+1-n}L_j  + \epsilon\frac{L_n^2 q_{\nu}}{c^2 x_{\nu}^2 }\,.\nonumber\\
&&
\ea
Using the separability lemma in the main text for $f_\nu={\mathcal{D}_{\nu}R_{\nu}}/({q_{\nu} R_{\nu}})$, the Lorenz condition \eqref{eq: appendix Lorenz condition}
 implies the following separated equations for
 the mode functions $R_{\nu}$:
\begin{equation}\label{eq: separation equations}
    \mathcal{D}_\nu R_{\nu} = q_\nu f_\nu R_{\nu}\,,
\end{equation}
where the polynomials $f_\nu$ are given by \eqref{sepfnu}, and are characterized by $(n-1)$ separation constants $C_j, j=0,\dots, n-2$.

It is now possible to show that the  Proca equation takes the form (valid both in odd and even dimensions)
\begin{equation}
    \nabla_b F^{ba} - m^2 \potential^a =  B^{ab} \nabla_{b} J\,,
\end{equation}
where the ``current" $J$ is given by
\begin{equation}
    J = \Box Z - 2 i \mu \xi_{a} B^{ab} \partial_{b}Z - m^2 Z\,.
\end{equation}
The explicit form of the box operator, $\Box$, in all dimensions is given in \cite{Sergyeyev:2007gf}. This yields the following formula for $J$:
\begin{equation}\label{eq: current J}
    J = Z \sum_{\nu=1}^{n}\frac{1}{U_{\nu}R_{\nu}}\left[ \mathcal{D}_{\nu} - m^2 \left( -x_{\nu}^2 \right)^{n-1} \right]R_{\nu}\,.
\end{equation}
Upon using the lemma again this amounts to
\be
 J = Z (\mu^2C_0-m^2)\sum_{\nu=1}^{n}\frac{(-x_\nu^2)^{n-1}}{U_{\nu}}=
 Z (\mu^2C_0-m^2)\,,
\ee
which vanishes if and only if we have
\be\label{C0}
C_0=\frac{m^2}{\mu^2}\,.
\ee

To summarize, the solution of the Proca equation \eqref{eq: Proca equation} in the general Kerr--NUT--AdS spacetimes in all dimensions can be found in the LFKK form  \eqref{LFKK3}, \eqref{sep3}, for the mode functions $R_\nu$ satisfying the ordinary differential equations \eqref{eq: separation equations}. Such a solution is general in that it is characterized by $(D-1)$ independent separation constants $\{\mu, C_1,\dots C_{n-2}, L_0,\dots, L_{n-1+\epsilon}\}$; the constant $C_0$ is not independent and is fixed by the mass of the vector according to equation~\eqref{C0}.
Moreover, in four dimensions all three massive polarizations can be extracted from this solution \cite{Dolan:2018dqv}. It remains to be seen whether the same remains true in a general dimension $D$, that is, whether all $D-1$ polarizations of the massive and $(D-2)$ polarizations of the massless vector field are encoded in this solution.


\providecommand{\href}[2]{#2}\begingroup\raggedright\endgroup

\end{document}